\documentclass[utf8]{frontiersSCNS} 
\usepackage{url,hyperref,lineno,microtype,subcaption}
\usepackage[onehalfspacing]{setspace}
\def\keyFont{\fontsize{8}{11}\helveticabold }
\def\firstAuthorLast{Onishi {et~al.}} 
\def\Authors{Seita Onishi\,$^{1}$, Ulrike Stockert\,$^{1}$, Seunghyun Khim\,$^{1}$, Jacintha Banda\,$^{1}$, Manuel Brando\,$^{1}$ and Elena Hassinger\,$^{1,2,*}$}
\def\Address{$^{1}$ Max Planck Institute for Chemical Physics of Solids, Dresden, Germany \\
$^{2}$ Physics Department, Technical University Munich, Germany}
\def\corrAuthor{Elena Hassinger}
\def\corrEmail{elena.hassinger@cpfs.mpg.de}
\begin{document}
\onecolumn
\firstpage{1}
\title[Low-Temperature Thermal Conductivity of CeRh$_{2}$As$_{2}$]{Low-Temperature Thermal Conductivity of the Two-Phase Superconductor CeRh$_{2}$As$_{2}$} 
\author[\firstAuthorLast ]{\Authors} 
\address{} 
\correspondance{} 
\extraAuth{}
\maketitle
\begin{abstract}
CeRh$_2$As$_2$ is a rare unconventional superconductor ($T_\mathrm{c}=0.26\,\mathrm{K}$) characterized by two adjacent superconducting phases for a magnetic field $H \parallel c$-axis of the tetragonal crystal structure. Antiferromagnetic order, quadrupole-density-wave order ($T_0 = 0.4\,\mathrm{K}$) and the proximity of this material to a quantum-critical point have also been reported: The coexistence of these phenomena with superconductivity is currently under discussion. Here, we present thermal conductivity and electrical resistivity measurements on a single crystal of CeRh$_2$As$_2$ between 60~mK and 200~K and in magnetic fields ($H \parallel c$) up to $8\,\mathrm{T}$. Our measurements at low $T$ verify the Wiedemann-Franz law within the error bars. The $T$ dependence of the thermal conductivity $\kappa(T)$ shows a pronounced drop below $T_\mathrm{c}$ which is also field dependent and thus interpreted as the signature of superconductivity. However, the large residual resistivity and the lack of sharp anomalies in $\kappa(T)$ at the expected transition temperatures clearly indicate that samples of much higher purity are required to gain more information about the superconducting gap structure. 
%


%
\tiny
 \keyFont{ \section{Keywords:} superconductivity, low temperature, heavy fermion, local inversion symmetry breaking, thermal conductivity, Wiedemann-Franz law} 
\end{abstract}

\section{Introduction}
CeRh$_2$As$_2$ is a newly discovered unconventional heavy-fermion superconductor with $T_\mathrm{c} = 0.26\,\mathrm{K}$~\citep{Khim2021}. Two different superconducting states occur when the magnetic field $H$ is applied along the crystallographic $c$-axis, as shown in the $H - T$ phase diagram in Fig.~\ref{kappa}A. The low-field superconducting state SC1 changes into to the high-field superconducting state SC2 at $\mu_0 H^\star \approx 4\,\mathrm{T}$. At this field the $T_\mathrm{c}$-vs-$H$ phase boundary line shows a sharp kink. Remarkably, the SC2 state has an extremely large upper critical field $\mu_0 H_\mathrm{c2} = 14\,\mathrm{T}$. For fields applied within the basal plane $H \perp c$ only the SC1 state exists and is Pauli limited with an upper critical field $\mu_0 H_\mathrm{c2} = 2\,\mathrm{T}$. This field exceeds the BCS Pauli limit by a factor of 4. The anisotropy of $H_\mathrm{c2}$ and the superconducting phase diagrams were explained by a model based on the crystal symmetry: The tetragonal CaBe$_2$Ge$_2$-type structure of CeRh$_2$As$_2$ is globally centrosymmetric but breaks inversion symmetry locally at the Ce site. As a consequence, Rashba spin-orbit coupling arises with alternating sign in the two adjacent Ce layers. In this picture, SC1 is identified as an even-parity superconducting state and SC2 as an odd-parity superconducting state~\citep{Khim2021,Cavanagh2022}. The angle dependence of the critical fields agrees with this interpretation~\citep{Landaeta2022}.

The behavior in the normal state is also unconventional and complex. It mainly originates from two key properties of CeRh$_2$As$_2$: its vicinity to a quantum-critical point and the unusual situation in which the energy difference between the two lowest-lying crystal-electric-field doublets of the Ce$^{3+}$ ions is comparable to the Kondo temperature ($T_\mathrm{K} \approx 30\,\mathrm{K}$) of this material. The first property is evidenced by the large increase of the specific heat, $C(T)/T \propto T^{-0.6}$, towards low temperatures, very low ordering temperatures and the presence of superconductivity \citep{Khim2021,Hafner2021}. Furthermore, a non-Fermi-liquid behavior is observed in resistivity, $(\rho - \rho_0) \propto \sqrt{T}$ ~\citep{Khim2021,Hafner2021}. The second property allows a substantial mixing of the excited crystal-electric-field states into the ground state by the Kondo interaction. Hence, quadrupolar degrees of freedom occur in the heavy bands at the Fermi level. Nesting can then promote a phase transition into a quadrupole-density-wave state which was proposed to exist below a temperature $T_0 \approx 0.4\,\mathrm{K}$~\citep{Hafner2021}. In addition, an increase of the As(2) nuclear quadrupole resonance line width below $T_\mathrm{N} \approx 0.3\,\mathrm{K}$ was interpreted as the onset of antiferromagnetic order~\citep{Kibune2021}. The influence of those phases on the superconducting states has still to be investigated thoroughly.

Many details of the normal-state and superconducting properties of CeRh$_2$As$_2$ are unsettled up to now. For example, little is known about the superconducting gap structure and the presence or absence of nodes. The only available information is from muon spin resonance experiments: The weak $T$ dependence of the superconducting relaxation rate points to a gap structure with point-like nodes or a full gap~\citep{Khim2022}. This motivated us to study the thermal conductivity $\kappa$ which depends sensitively on the available heat carriers and on the relevant scattering processes. In the past, measurements of $\kappa$ have been applied successfully to study the gap structure of unconventional superconductors \citep{Izawa05, Dong10, Shakeripour10, Reid12}. So, this technique might be able to reveal differences in the gap structure of the two superconducting states in CeRh$_2$As$_2$. Another fundamental relation that can be verified by these experiments is the Wiedemann-Franz law. A violation of this law in the $T = 0$ limit would provide a direct proof for the breakdown of the quasiparticles picture, as it has been proposed for systems at a quantum-critical point~\citep{Tanatar07,Podolsky07,Kim09,YRS-12-1,Taupin2015}. In order to obtain a comprehensive picture about the heat transport in CeRh$_2$As$_2$, we measured $\kappa$ in a large $T$ range, between 60~mK and 200~K, which covers $T_\mathrm{c}$, $T_0$ as well as $T_\mathrm{K}$. At low temperatures, we also studied the magnetic-field dependence with fields $H \parallel c$ up to 8\,T across the transition field $H^{\star}$ between SC1 and SC2. We supplement these investigations by electrical resistivity measurements and a comparison of our data with the Sommerfeld coefficient $C/T (T \rightarrow 0)$.
%
%

Our measurements reveal no clear anomalies in the thermal conductivity as a function of temperature at the phase transition temperatures $T_\mathrm{c}$ and $T_0$, probably due to scattering from defects in our sample. A crude extrapolation to $T = 0$ of the Lorenz ratio from above 0.4\,K confirms the validity of the Wiedemann-Franz law within the error bars. Superconductivity induces a pronounced drop of $\kappa(T)/T$ below $T_\mathrm{c}$ that shifts with magnetic field as expected from the behavior of $H_\mathrm{c2}$. However, better samples are needed to obtain clear information about the gap structure.
\section{Methods and Experimental Details}
We measured the low-temperature thermal conductivity $\kappa (T)$ and the electrical resistivity $\rho (T)$ of single-crystalline CeRh$_2$As$_2$ between $T = 60\,\mathrm{mK}$ and $1.2\,\mathrm{K}$ and for magnetic fields up to $8\,\mathrm{T}$. Both, the heat current and the electrical current were applied within the $ab$-plane of the crystal, while the magnetic field was applied parallel to the $c$ axis. A second sample was used to extend zero-field measurements to about $300\,\mathrm{K}$. Both samples had dimensions of approximately $2 \times 0.5 \times 0.1$ mm$^3$. 

Low-temperature measurements were performed in a dilution refrigerator using a home-made setup. The thermal conductivity was measured by a four-point steady-state method using two RuO$_x$ thermometers and a resistive heater. The sample thermometers were calibrated in-situ for each temperature sweep against a field-calibrated thermometer at the cold bath. The field was always changed at high temperature in the normal state of the sample. The electrical resistivity was determined by an ac technique. $\kappa$ and $\rho$ were obtained in subsequent measurements using the same contacts. 

High-$T$ measurements were performed in a commercial Physical Property Measurement System (PPMS from Quantum Design) using a modified sample holder for the thermal transport option suitable for small single crystals (see \citep{Stockert2017} which contains a photo of part of the setup). Due to the considerable contact width compared to the sample dimensions, the uncertainty of the geometry factor of the high-temperature measurement is relatively large. Instead, the geometry factor chosen for sample 2 in Fig. \ref{kappa}B was determined by matching low- and high-$T$ $\kappa$ and $\rho$ data. It agrees with the geometry factor from sample size and contact separation within the uncertainty, i.e. a difference of 20\,\% between the two geometry factors. For $T > 200$~K, the measured thermal conductivity strongly increases due to thermal radiation losses. Therefore, we neither show nor discuss $\kappa$-data in this $T$ range.
%
%
\section{Results}
\subsection{Zero-field thermal and electrical transport between $2\,\mathrm{K}$ and $200\,\mathrm{K}$}
\begin{figure}[tb]
\begin{center}
\includegraphics[width=1.0\textwidth]{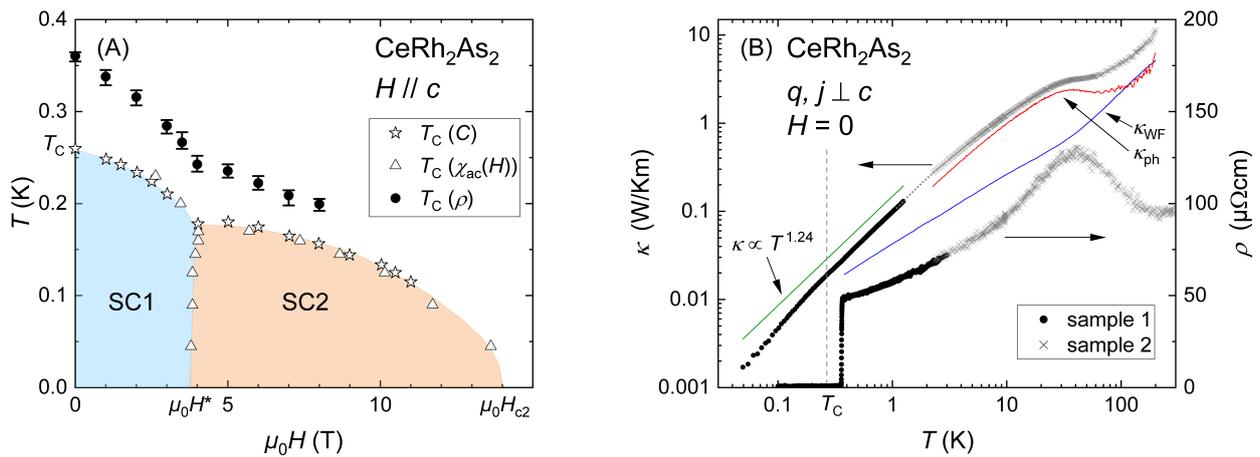}
\end{center}
\caption{(Color online) (A) Superconducting phase diagram of CeRh$_2$As$_2$ for $H // c$. The values of $T_\mathrm{c}$ at the phase boundaries estimated from specific-heat and field-dependent ac susceptibility measurements are reproduced from Ref.~\cite{Khim2021}. In addition, we show $T_\mathrm{c}$ determined from our resistivity measurements as the temperature at which $\rho$ drops to 50\% of the normal-state value. Error bars indicate the transition width. (B) Temperature dependence of the thermal conductivity $\kappa(T)$ and of the electrical resistivity $\rho (T)$ of CeRh$_{2}$As$_2$. The low-$T$ and high-$T$ data have been measured on two different samples using two different setups. The dotted line, which connects the high-$T$ and low-$T$ $\kappa(T)$ data, is a guide to the eye. The bulk $T_\mathrm{c}$ estimated from specific-heat measurements is marked by a dashed vertical line. Estimates of the electron and phonon contributions to the thermal conductivity, determined as explained in the text, are shown. The green line emphasizes that $\kappa \propto T^{1.24}$ between $1.2\,\mathrm{K}$ and $T_\mathrm{c}$.}
\label{kappa}
\end{figure}

Fig.~\ref{kappa}B shows the thermal conductivity of CeRh$_2$As$_2$ between $200\,\mathrm{K}$ and $60\,\mathrm{mK}$. For comparison, we also plot the electrical resistivity $\rho(T)$ between $300\,\mathrm{K}$ and $100\,\mathrm{mK}$, which reflects the behavior previously observed in other samples~\citep{Khim2021}. In this section, we only focus on data above $2\,\mathrm{K}$. $\kappa(T)$ decreases monotonously with decreasing temperature. A plateau is reached at around $40\,\mathrm{K}$. On the contrary, the resistivity $\rho(T)$ increases with decreasing temperature due to Kondo scattering. At $T_{K} \approx 30\,\mathrm{K}$ a maximum is reached. Towards lower $T$, $\rho(T)$ decreases due to the onset of coherence and thermal depopulation of the excited crystal-electric-field levels. From these data it is not directly clear whether the plateau in $\kappa(T)$ is related to the maximum in $\rho(T)$ or not. We can, however, consider that the thermal conductivity of standard nonmagnetic metals is the sum of two contributions, namely that of phonons ($\kappa _{\mathrm{ph}}$) and that of charge carriers ($\kappa _{\mathrm{el}}$). A rough estimate of $\kappa _{\mathrm{el}}$ can be obtained from the electrical resistivity by simply assuming that the Wiedemann-Franz law is valid. In this case, $\kappa _{\mathrm{el}} \approx \kappa _{\mathrm{WF}} = L_0 T/\rho$ with the Lorenz constant $L_0 = \frac{\pi^2}{3}(\frac{k_\mathrm{B}}{e})^2 = 2.44 \times 10^{-8}$ V$^2 /$ K$^2$. By subtracting $\kappa _{\mathrm{WF}}$ from the total thermal conductivity we obtain an estimate for $\kappa _{\mathrm{ph}}$. Both contributions to $\kappa$ are shown in Fig.~\ref{kappa}B. It can be clearly seen, that in the temperature range between roughly $2\,\mathrm{K}$ and $200\,\mathrm{K}$, the thermal conductivity is dominated by the phonon contribution. The uncertainty in our estimate of $\kappa _{\mathrm{el}}$ from $\rho(T)$ is therefore not relevant for the qualitative behavior of $\kappa _{\mathrm{ph}}$. In particular, $\kappa _{\mathrm{ph}}$ is definitely responsible for the plateau at $40\,\mathrm{K}$. 

Qualitatively, the temperature dependence of $\kappa _{\mathrm{ph}}$ can be understood as follows: Starting from low $T$, $\kappa _{\mathrm{ph}}$ of clean materials initially increases with increasing phonon density. Upon further heating, phonon-phonon interactions become more relevant with allowed Umklapp scattering, and $\kappa_{\mathrm{ph}}$ starts to decrease again. Therefore, $\kappa _{\mathrm{ph}}$ usually exhibits a maximum at intermediate temperatures. The exact temperature dependence of $\kappa _{\mathrm{ph}}$ as well as the magnitude and position of the maximum depend on the relevant scattering processes and the purity of the material. Our crystals exhibit rather large residual resistivities with $\rho(300\,\mathrm{K}) / \rho(0.4\,\mathrm{K}) \approx 2$ indicative of a considerable number of scattering centers. This may be the reason for the absence of a more pronounced maximum in $\kappa _{\mathrm{ph}} (T)$. A weak $T$ dependence of $\kappa _{\mathrm{ph}}$ without pronounced maximum is typical of Ce-based heavy-fermion compounds, and has been observed, e.g. in CeCu$_4$Al~\citep{TB-11-4} and in CeRh$_2$Ga$_2$~\citep{TB-17-1}. 

Between $2\,\mathrm{K}$ and $10\,\mathrm{K}$, in CeRh$_{2}$As$_{2}$, the inelastic scattering of charge carriers below the Kondo coherence maximum and the non-Fermi-liquid behavior observed at low-$T$ lead to a large uncertainty of estimating the phonon thermal conductivity via the Wiedemann-Franz law. Lacking a better method, we nevertheless continue in the same way. The estimated phonon thermal conductivity of CeRh$_2$As$_2$ in that temperature range exhibits roughly a linear temperature dependence, $\kappa(T) _{\mathrm{ph}} \propto T$, which may be somewhat surprising. In a very simple picture $\kappa _{\mathrm{ph}}$ is expected to follow a $T^3$-function. This type of behavior can be derived within the Boltzmann formalism assuming a momentum and $T$-independent phonon mean free path and velocity. However, the situation in real materials is much more complicated. Different types of scattering mechanisms lead to other temperature dependencies. For instance, scattering of phonons from electrons and dislocations may result in $\kappa _{\mathrm{ph}} \propto T^2$ (\cite{Ziman}) and phonon specular reflection results in an exponent between 2 and 3~\citep{Li2008}. A linear-in-$T$ dependence of $\kappa _{\mathrm{ph}}$ has also been found and ascribed to the existence of dislocation networks~\citep{Kapoor74,TB-78-1}. A quasi-linear-in-$T$ behavior was found in the above mentioned Ce-based heavy-fermion compounds CeCu$_4$Al~\citep{TB-11-4} and CeRh$_2$Ga$_2$~\citep{TB-17-1}.  
\subsection{Zero-field thermal and electrical transport below $2\,\mathrm{K}$}
\begin{figure}[tb]
\begin{center}
\includegraphics[width=1.0\textwidth]{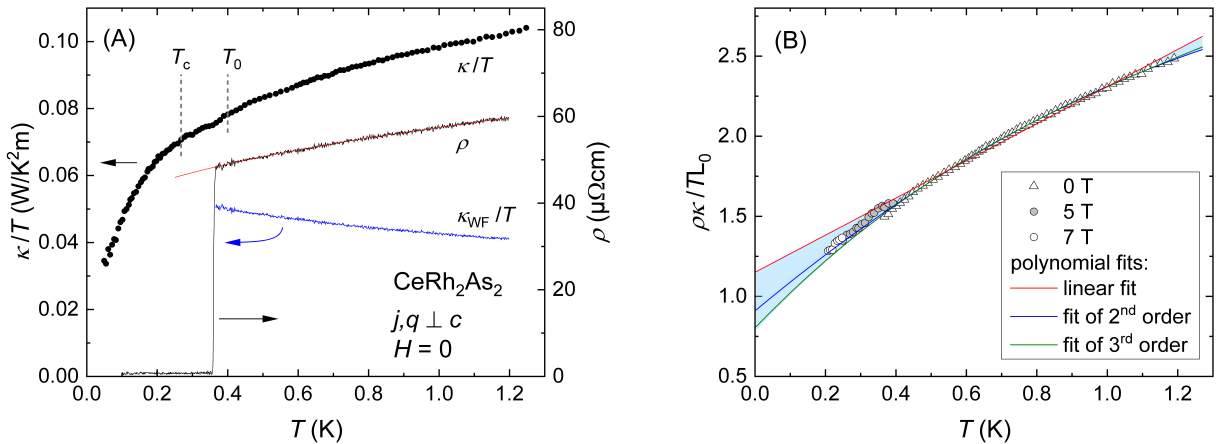}
\end{center}
\caption{(Color online) (A) $T$ dependence of the thermal conductivity of CeRh$_{2}$As$_2$ plotted as $\kappa(T)/T$ vs $T$ and compared to electrical resistivity $\rho(T)$ measured using the same contacts. A fit $\rho =\rho_0 + A\sqrt{T}$ for $T > 0.5$~K is shown as a red line. The electronic contribution to the thermal conductivity estimated from $\rho(T)$ using the Wiedemann-Franz law is plotted as $\kappa(T)_{\mathrm{WF}}/T$. (B) $T$ dependence of the Lorenz ratio normalized to $L_0$. The lines are fits to the zero-field data. The shaded area indicates the range of fit reliability as explained in the text.}
\label{WF}
\end{figure}
Now we take a look at the low-$T$ behavior of $\rho(T)$ and $\kappa(T)$ in zero magnetic field summarized in Fig.~\ref{WF}A. The resistivity shows the superconducting transition at $0.36\,\mathrm{K}$. This temperature is larger than the bulk $T_\mathrm{c} = 0.26\,\mathrm{K}$ determined from specific heat, as shown in Fig.~\ref{kappa}A. This is probably due to defect-mediated percolation superconductivity~\citep{Khim2021}. Importantly, Fig.~\ref{kappa}A demonstrates how the suppression of the superconducting transition in $\rho(T)$ with magnetic field follows the behavior of the bulk $T_\mathrm{c}$, thus excluding the possibility of a superconducting impurity phase. The low-temperature normal-state resistivity exhibits a rather unusual temperature dependence, $\rho(T) \propto \sqrt{T}$~\citep{Hafner2021}. In simple metals $\rho$ can be treated as a sum of three contributions: $\rho = \rho_0 + \rho_{\mathrm{el-ph}} + \rho_{el-el}$. $\rho_0$ is a constant residual value due to scattering of charge carriers by impurities, which dominates in the zero-$T$ limit. Scattering of electrons by phonons gives rise to a $\rho_{\mathrm{el-ph}} \propto T^5$ increase. Electron-electron scattering results in the fomation of a Fermi liquid with $\rho_{el-el} = AT^{2}$ for $T \rightarrow 0$. In heavy-fermion metals, electron-electron scattering adds a considerable contribution since the electron mass is renormalized to a larger effective mass $m^{*}$ and $A \propto (m^{*})^{2}$~\citep{Kadowaki1986}. Deviation from Fermi-liquid behavior is primarily found in systems located at the quantum-critical point~\citep{Custers2003}. This could indeed be the case of CeRh$_2$As$_2$ in which our fit with $\rho = \rho_0 + A\sqrt{T}$ shows a good match for $T > 0.5\,\mathrm{K}$ (red line in Fig.~\ref{WF}A). However, the origin of this unusual $T$ dependence is so far not clear. Since it is found in a limited temperature range, it might also be a crossover region between different behaviors, e.g., the Kondo coherence region below $30\,\mathrm{K}$ and a Fermi-liquid region at low-$T$. In addition, an upturn in $\rho(T)$ below $T_0$ was previously observed~(\cite{Hafner2021}). This was attributed to the opening of a gap at the Fermi surface due to nesting at $T_0$. This feature is not resolved in our measurement, possibly due to strong impurity scattering or because it appears across the onset of superconductivity at $0.36\,\mathrm{K}$. In the zero-temperature limit, a Fermi-liquid $T^2$ dependence should be recovered if the system is not right at a quantum-critical point. From the observed $T$ dependence, it is not possible to determine a reasonable value for $\rho_0$ from these data.


The thermal conductivity changes smoothly with decreasing temperature and follows a power-law behavior $\kappa(T) \propto T^{n}$ with $n = 1.24$ below $1.2\,\mathrm{K}$ (see green line in Fig.~\ref{kappa}B), but no sharp anomalies could be detected neither at $T_0$ nor at $T_\mathrm{c}$. Above $T_0$ we observe a weakly concave $T$ dependence. This is unusual and probably due to a combination of both the unconventional temperature dependencies of electron and phonon contributions. Therefore, we cannot subtract the phonon contribution and extract the value of $\kappa/T$ of the normal state in the zero-temperature limit. Lowering the temperature further, $\kappa/T$ bends down at approximately $0.2\,\mathrm{K}$ and shows a steep slope of the data at lowest $T$. We attribute this downward bending of $\kappa /T$ to the onset of superconductivity. Such behavior is not unusual. From theory, it is difficult to predict the effect of superconductivity on thermal transport. In general, the inhanced quasiparticle scattering time below $T_\mathrm{c}$ competes with the suppression of the quasiparticle population, and both an increase and a drop of $\kappa$ have been observed when cooling below $T_\mathrm{c}$ . In all cases, $\kappa$ goes to zero in the zero-temperature limit, but $\kappa/T$ can stay finite in the superconducting state if the gap structure has nodes. The downward bending in CeRh$_2$As$_2$ is then due to the reduced number of heat carriers towards low temperature since Cooper pairs do not carry heat. Our field-dependent measurements presented below provide further confirmation, that the bending in $\kappa/T$ is caused by the onset of superconductivity. 
\subsection{ Wiedemann-Franz law}
In the temperature range below $2\,\mathrm{K}$, the electronic contribution to the thermal conductivity becomes increasingly relevant (see a comparison of $\kappa/T$ and $\kappa_\mathrm{WF}/T$ in Fig. \ref{WF}A). Normally, the Wiedermann-Franz law is studied and established for the electronic contribution to thermal conductivity, i.e. $L = \rho \kappa_{el} / T$, which can tell us if scattering processes have different effects on thermal and electrical transport of charge carriers~\citep{Jaoui2021}. In the zero-temperature limit $L = L_0$ is expected in a normal metal where the Wiedemann-Franz law should be fulfilled, but might be violated near a quantum-critical point \citep{Tanatar07,Podolsky07,Kim09,YRS-12-1,Taupin2015}. However, as described above, we are not able to determine and remove $\kappa_{ph}$ from $\kappa$ to extract the pure electronic contribution. 
Therefore, we use a different approach. We calculate the Lorenz ratio $L = \rho \kappa / T$ taking the total thermal conductivity and normalize it by the Lorenz constant $L_0$. This ratio is also expected to reach one in the zero-temperature limit if the Wiedemann-Franz law is fulfilled, since $\kappa_{ph}$ goes to zero.


Fig.~\ref{WF}B shows the normalized Lorenz ratio $L/L_0$ for CeRh$_2$As$_2$ in the $T$ range below $1.2\,\mathrm{K}$. $L/L_0$ is well above one but decreases towards one with decreasing $T$. Most likely, the phonon contribution to $\kappa$ is still considerable in this temperature range in our material. Due to the onset of superconductivity, our calculation of $L/L_0$ in zero field is limited to temperatures above $0.36\,\mathrm{K}$. These data mainly reflect the behavior of CeRh$_2$As$_2$ above $T_0$, but an influence of the transition at $T_0 = 0.4$~K cannot be excluded. Therefore, and because of the unusual temperature dependencies of $\kappa$ and $\rho$, any extrapolation of the data to $0\,\mathrm{K}$ goes along with large uncertainties. As rough attempts we show polynomial fits up to the $3^{\mathrm{rd}}$ order for $T > 0.4\,\mathrm{K}$, without any physical reasoning. A linear fit does not describe the data well and extrapolates to $L/L_0 = 1.2$. Polynoms of higher order better fit the data and extrapolate to values of $L/L_0 < 1$. The corresponding fits are shown as lines Fig.~\ref{WF}B. Using only part of the data range for fitting slightly changes the results. Nevertheless, as long as we use at least the data up to 1~K, all fits fall in the range shaded in blue in Fig.~\ref{WF}B, which is roughly limited by the linear fit and the polynom of $3^{\mathrm{rd}}$ order on the whole data set below $1.2\,\mathrm{K}$. Within the error bars of our analysis, $L/L_0$ extrapolates to one at $T \rightarrow 0$ as expected for a Fermi liquid. However, moderate deviations from this ideal value cannot be excluded due to the limited $T$ range for the data extrapolation. 

A more accurate evaluation of $L/L_0$ in the unordered state might be achieved in magnetic fields if we assume that the only effect of the magnetic field is to suppress both $T_0$ and $T_\mathrm{c}$ without changing the paramagnetic state. Therefore, we also show $L/L_0$ in 5 T and 7 T at lower $T$ for comparison in Fig.~\ref{WF}B. These data points follow the extrapolation of the zero-field curve.
%
%
\subsection{Magnetic field dependence}
%
%
\begin{figure}[tb]
\begin{center}
\includegraphics[width=1.0 \textwidth]{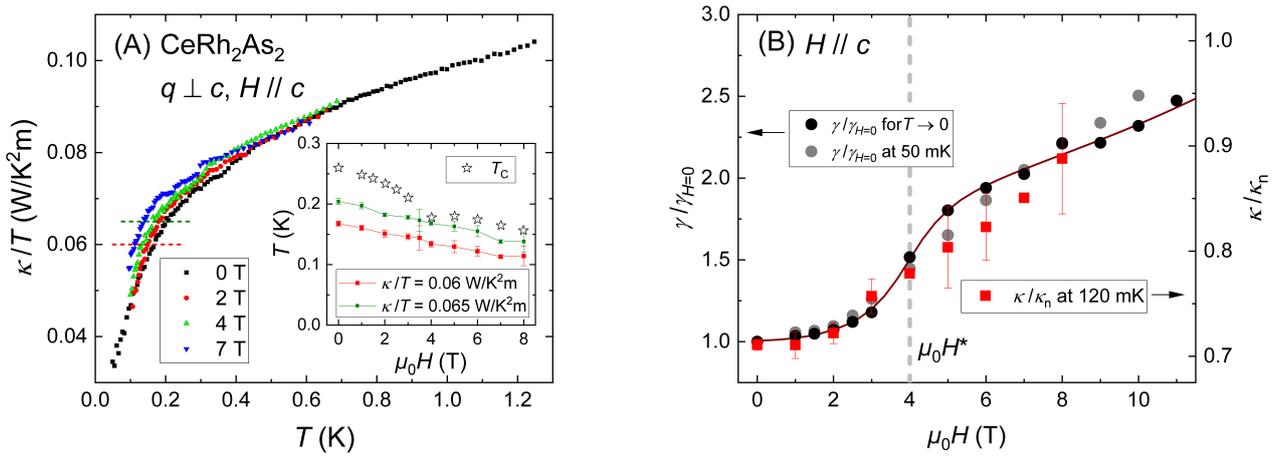}
\end{center}
\caption{(Color online) (A) Main plot: Thermal conductivity of CeRh$_{2}$As$_2$ divided by temperature in various magnetic fields. The dashed lines indicate cuts at constant $\kappa /T$, which are used for the inset. Inset: The stars indicate the field dependence of $T_\mathrm{c}$ taken from Ref.~\cite{Khim2021} and are compared with points taken from our data at constant $\kappa /T$. (B) Field dependence of the thermal conductivity at $120\,\mathrm{mK}$ normalized to the value in the normal state, $\kappa / \kappa_n$. For comparison, the field dependence of the specific heat coefficient $\gamma = C/T$ taken at constant $T = 50\,\mathrm{mK}$ and extrapolated for $T \rightarrow 0$, normalized to the value at $H = 0$, is plotted. The continuous line is a guide to the eye. The dashed perpendicular line marks the transition field $\mu_0 H^\star$ between SC1 and SC2.}
\label{3and4}
\end{figure}
In order to clarify whether the low-temperature downward bending of $\kappa /T$ at roughly 0.2\,K is related to superconductivity, we studied the influence of magnetic fields $H$ up to $8\,\mathrm{T}$ parallel to the $c$-axis. Fig.~\ref{3and4}A shows $\kappa(T)/T$ in various fields. Above $0.5\,\mathrm{K}$ we find an almost field-independent behavior. The change of $\kappa /T$ with field is less than 3\% and below the resolution of our measurement. This negligible field dependence is consistent with the weak magnetoresistance for charge transport~\citep{Hafner2021}. No field dependence is expected from phonon transport, as well. Below about $0.4\,\mathrm{K}$ ($\approx T_0$), $\kappa$ increases slightly with field. Although a clear anomaly in the zero-field curve is missing, this increase suggests that the order at $T_0$ suppresses $\kappa$. This is in agreement with an increase in the electrical resistivity observed in small in-plane magnetic fields that was explained by a reconstruction of the Fermi surface due to nesting~\citep{Hafner2021}. When this order disappears with fields along the $c$-axis, $\kappa$ increases and $\rho$ decreases.

At lower $T \approx 0.2\,\mathrm{K}$ the downward bending of $\kappa /T$ is shifted to lower temperature with increasing $H$. This is accompanied by increasing values of $\kappa/T$ at constant $T$. Although the bending of $\kappa /T$ is clearly visible, it is not possible to define a characteristic onset temperature or a $T_\mathrm{c}$ from our data directly. Therefore, we use the field dependence of constant $\kappa/T$ to quantify the influence of $H$. In the inset of Fig.~\ref{3and4}A we plot the temperatures $T (\kappa /T = \mathrm{const})$ at which $\kappa /T$ reaches $0.06\,\mathrm{W/K^2m}$ and $0.065\,\mathrm{W/K^2m}$. The corresponding values are indicated by dashed lines in the main plot of Fig.~\ref{3and4}A. For comparison, we also plot the superconducting transition temperature $T_\mathrm{c}$ from Ref.~\citep{Khim2021} determined by specific-heat measurements. The inset of Fig.~\ref{3and4}A clearly demonstrates that the downward bending of $\kappa /T$ moves to lower temperature with increasing magnetic field. The field dependence of $T (\kappa /T = \mathrm{const})$ roughly follows that of $T_\mathrm{c}$. However, the characteristic behavior of $T_\mathrm{c}(H)$ with a kink at the transition between the two superconducting phases at $\mu_0 H^\star = 4\,\mathrm{T}$ is not well reproduced. This is in contrast to our resistivity results on the same sample that show a clear kink at $T^{\star}$, see Fig.~\ref{kappa}A. The discrepancy between electrical and thermal conductivity may be due to the limited resolution of our analysis, which does not reveal any feature in $T (\kappa /T = \mathrm{const})$ vs. $H$.

In summary, the impact of superconductivity on $\kappa (T)$ of CeRh$_2$As$_2$ in our experiments is weak, probably due to residual states which contribute to thermal transport, despite the condensation of quasiparticles into the superconducting state. In the trivial case, such states are defects, which cause the presence of normal quasiparticles. A systematic study on Sr$_2$RuO$_4$ shows a broadening of the superconducting transition in $\kappa (T)$ with increasing impurity concentration~\citep{Suzuki02}. A similar trend is seen in type-I, elemental superconductors~\citep{Hulm49}. In CeIrIn$_5$, a heavy-fermion superconductor, $\kappa/T$ ($T< T_\mathrm{c}$) shows a peak. With disorder, the peak disappears and becomes a downward bend from the normal-state value at $T_\mathrm{c}$~\citep{Shakeripour10}. So, measurements on CeRh$_2$As$_2$ samples of higher purity are definitively required to study the influence of superconductivity on $\kappa (T)$ in more detail.

The temperature dependence of $\kappa(T)$ in the temperature range below $T_\mathrm{c}$ provides information about the gap structure. For instance, an exponential $T$ dependence indicates a full superconducting gap whereas a power-law-behavior points to the presence of nodes. As a consequence, the extrapolation of the thermal conductivity $\kappa/T$ towards $T = 0$ 
will be finite if the superconducting gap has nodes, but it will be zero if the gap is finite everywhere on the Fermi surface~\citep{Shakeripour10}. \\The magnetic field dependence also depends on the superconducting gap structure. The field dependence of thermal conductivity and specific heat in a superconductor in the zero-temperature limit is comparable if scattering of the quasiparticles with vortices can be neglected. Both quantities are sensitive to the increase of the density of states of the normal states inside the vortex core - resulting in a linear increase in field - and to the Volovik effect~\citep{Volovik1993} which can lead to different field dependencies depending on the pairing state~\citep{Bang2010}, for example the famous $\sqrt{H}$ dependence for line nodes. On the other hand, scattering with vortices can lead to a decrease in $\kappa$ with increasing field that does not affect the specific heat. Unfortunately, the extrapolation of $\kappa$ to $T = 0$ does not make sense for our data since we are at 0.25~$T_\mathrm{c}$ for $H = 0$ and approach 0.6~$T_\mathrm{c}$ at $8\,\mathrm{T}$, all not low enough. Therefore, we only report on the observed field dependence without a detailed discussion of the physics.
With that in mind, we investigate the field dependence of $\kappa$ in the superconducting state at $120\,\mathrm{mK}$, which was the lowest temperature at which $\kappa$ was determined for all fields. We have normalized $\kappa$ to a field-independent estimate of the normal-state value at that temperature, $\kappa_\mathrm{n}\,(120 \mathrm{mK}) = 8.7 \times 10^{-3}\,\mathrm{W/Km}$ determined from a linear extrapolation of the temperature dependence of $\kappa/T$ in high magnetic fields ($\mu_0 H \geq 7\,\mathrm{T}$, $T > 0.3\,\mathrm{K}$). Using zero-field data between $0.4\,\mathrm{K}$ and $1\,\mathrm{K}$ instead for the estimation of $\kappa_\mathrm{n}$, it yields a similar value of $8.5 \times 10^{-3}\,\mathrm{W/Km}$. The results are plotted in Fig.~\ref{3and4}B and compared to the field dependence of the specific-heat coefficient $\gamma = C/T$ taken at constant $T = 50\,\mathrm{mK}$ and extrapolated for $T \rightarrow 0$, normalized to the value at $H = 0$. Note that there is a large residual $\gamma$ even in the $T\rightarrow 0$ limit~\citep{Khim2021}, which changes with the observed $T_\mathrm{c}$~\citep{Brando2022}. This is an indication that the low-temperature $C/T$ value might be dominated by impurities. 

At low magnetic fields, $\gamma / \gamma_{H = 0}$ increases very slowly with field, a behavior typical of $s$-wave superconductivity~\citep{Shakeripour10}. This is in agreement with results of the penetration depth obtained from muon-spin-resonance experiments, where a nearly-full-gap temperature dependence was observed~\citep{Khim2022}. Around $H^\star$, a stronger increase occurs, followed by another region with weak, almost linear field dependence. The overall behavior for data points at $50\,\mathrm{mK}$ and in the zero-$T$ limit is similar, with a slightly steeper increase for $T \rightarrow 0$ around $H^\star$. The field dependence of $\kappa$ roughly follows the one of $\gamma$ with a small slope at low $H$ and a larger one around $H^\star$ when going from SC1 to SC2. Measurements at higher fields and lower $T$ are required to confirm that this change in slope is really due to the transition between the two superconducting phases. 
\section{Summary and Conclusion}
In this manuscript we have presented the thermal conductivity and electrical resistivity of single-crystalline CeRh$_2$As$_2$ between $60\,\mathrm{mK}$ and $200\,\mathrm{K}$. The high-temperature behavior is typical for a Ce-based Kondo-lattice system with the thermal conductivity dominated by phonons and the electrical resistivity governed by Kondo scattering and crystal-electric-field excitations. By contrast, the temperature dependence between $60\,\mathrm{mK}$ and $1.2\,\mathrm{K}$ of both quantities does not follow the expected behavior for a Fermi liquid. Within the uncertainties of our data and analysis, the data fulfill the Wiedemann-Franz law in the $T = 0$ limit. No sharp anomalies are detected by the thermal conductivity at the expected phase transition temperatures, $T_\mathrm{c}$ and $T_0$, seen in thermodynamic and resistivity measurements, but a downward bending of $\kappa/T$ at low temperature appears to be due to superconductivity. In fact, as expected, this feature shifts to lower temperatures when magnetic fields are applied along the $c$-axis. It seems that impurities in our sample and associated scattering hide the putative sharp features expected for both phase transitions. Hence, better sample quality is required to obtain clear information about the superconducting gap structure via thermal conductivity.

\section*{Conflict of Interest Statement}

The authors declare that the research was conducted in the absence of any commercial or financial relationships that could be construed as a potential conflict of interest.

\section*{Author Contributions}
SK grew the samples. US carried out the high-temperature measurements and the low-temperature field calibration. SO carried out the low-temperature experiments. US and SO analyzed the data with input from EH. JB and MB measured and analysed the specific heat. US and EH wrote the manuscript with input from all coauthors.

\section*{Funding}
We acknowledge funding from the Physics of Quantum Materials department and the research group “Physics of Unconventional Metals and Superconductors (PUMAS)” by the Max Planck Society. C.G. and E.H. acknowledge support from the German Science Foundation (DFG) through grant GE 602/4-1 Fermi-NESt. The research environment in Dresden benefits from the DFG Excellence Cluster Complexity and Topology in Quantum Matter (ct.qmat).

\section*{Acknowledgments}
We thank Christoph Geibel and Louis Taillefer for useful comments about the manuscript.

\bibliographystyle{frontiersinSCNS_ENG_HUMS} 

\end{document}